\documentclass[preprint]{elsarticle}
\usepackage{amsmath,amsfonts,amssymb,bm}
\numberwithin{equation}{section}
\begin{document}
\title{Decomposition of the Feynman kernel \\for a particle in a box}
\author{Seiji Sakoda}
\ead{sakoda@nda.ac.jp}
\address{Department of Applied Physics, National Defense Academy,
Hashirimizu\\
Yokosuka city, Kanagawa 239-8686, Japan}

\date{\today}

\begin{abstract}
We study the decomposition of the Feynman kernel for a particle in a box with $1/\sin^{2}\theta$ potential to find that the wellknown phase factor $-1$, which is correct for the case of the free particle, for reflection at boundaries  should be generalized depending on the parameter of the potential.
\end{abstract}

\maketitle

\section{Introduction}
The Feynman kernel for the free particle on a circle\cite{Schulman1968,SchulmanBook,KleinertBook} and for the one in a one dimensional box\cite{JankeKleinert,KleinertBook} are wellknown examples of solvable path integrals as a sum over classical paths by taking the topology of the system into account. The topology of a box appears as a phase factor $-1$ at each reflection by boundaries while the winding number of the path is entering as phase for the free particle on a circle.
On the other hand, in Schr\"{o}dinger picture of quantum mechanics, there exists a simple and exactly solvable model\cite{CKSbook} with a potential proportional to $1/\sin^{2}\theta$ ($\theta=\pi x/L$) on the interval $0<x<L$. This model is also known to be exactly solvable in Heisenberg picture thanks to the existence of the sinusoidal coordinate\cite{Odake_Sasaki}. 

We aim in this note to make it clear whether the phase factor $-1$ above can be universal and independent of the potential. To achieve this, we convert the eigenfunction expansion of the Feynman kernel for the Hamiltonian with $1/\sin^{2}\theta$ potential into the form of a sum over paths with difference in the number of reflections.

\section{Feynman kernel for the $1/\sin^{2}\theta$ potential}
Stationary Schr\"{o}dinger equation
\begin{equation}
\label{eq:Schrodinger}
	{\mathcal R}\left\{-\frac{\ d^{2}\ \ }{\ d\theta^{2}\ }+
	\frac{\ \nu(\nu-1)\ }{\ \sin^{2}\theta\ }\right\}\psi(\theta)=E\psi(\theta),
	\quad{\mathcal R}\equiv
	\frac{1}{\ 2m\ }\left(\frac{\ \pi\hbar\ }{L}\right)^{2}
\end{equation}
can be solved by the eigenfunction
\begin{equation}
\label{eq:eigenfunct}
	\phi^{(\nu)}_{n}(\theta)=2^{\nu}\varGamma(\nu)
	\sqrt{\frac{\ (n+\nu)n!\ }{\ 2\pi\varGamma(n+2\nu)\ }\,}
	\sin^{\nu}\theta
	C^{\nu}_{n}(\cos\theta)
\end{equation}
to yield the $n$-th eigenvalue $E^{(\nu)}_{n}=(n+\nu)^{2}{\mathcal R}$ for $n=0,\,1,\,2,\,\dots$. Here the parameter $\nu$ is assumed to be $\nu\ge1/2$ and  $C^{\nu}_{n}(\cos\theta)$ describes the Gegenbauer polynomial of the $n$-th order.

The stationary Scr\"{o}dinger equation defines a Hamiltonian
\begin{equation}
	H^{(\nu)}=\frac{\ p^{2}\ }{\ 2m\ }+
	\frac{\ \nu(\nu-1){\mathcal R}\ }{\ \sin^{2}\theta\ }
\end{equation}
where $p^{2}$ acts as $-\hbar^{2}d^{2}/dx^{2}$ on wavefunctions.
Since we know the complete set of eigenfunctions of this Hamiltonian,
we can immediately write out the eigenfunction expansion of the Feynman kernel as
\begin{equation}
\label{eq:kernel01}
	{\mathcal K}^{(\nu)}(\theta_{a},\theta_{b};\beta)=
	\sum_{n=0}^{\infty}e^{-\beta E^{(\nu)}_{n}/\hbar}
	\phi^{(\nu)}_{n}(\theta_{a})\phi^{(\nu)}_{n}(\theta_{b})
\end{equation}
for the Euclidean time evolution operator $e^{-\beta H^{(\nu)}/\hbar}$.
Note that this kernel is normalized to be suitable for integration with respect to $\theta_{a(b)}=\pi x_{a(b)}/L$ instead of $x_{a(b)}$.
(The derivation of this kernel by means of the path integral technique is given in Ch. 8.8 of ref.\cite{KleinertBook} but the method shown there does not treat the direct formulation of the path integral for the Hamiltonian above.)
By setting $\nu=1$, we find that $H^{(\nu)}$ reduces to the free particle in a box. It will be, therefore, interesting to find a suitable decomposition of this kernel into paths different in the number of reflections by boundaries for arbitrary values of $\nu$. 

Let us begin with a short time kernel ${\mathcal K}^{(\nu)}(\theta,\theta{'};\epsilon)$ for infinitesimally small $\epsilon$.
We will set $\lambda=2\epsilon{\mathcal R}/\hbar$ to write $\epsilon E^{(\nu)}_{n}/\hbar$ as $\lambda(n+\nu)^{2}/2$ in the following for the sake of simplicity. Then, by recalling the asymptotic expression for the modified Bessel function
\begin{equation}
	I_{\nu+n}\left(\frac{\ 1\ }{\ \lambda\ }\right)\sim
	\sqrt{\frac{\lambda}{\ 2\pi\ }\,}
	\exp\left[\frac{\ 1\ }{\ \lambda\ }-
	\frac{\ \{4(n+\nu)^{2}-1\}\lambda\ }{8}\right]
\end{equation}
for infinitesimally small positive $\lambda$, we observe
\begin{equation}
	e^{-\lambda(n+\nu)^{2}/2}\sim
	\sqrt{\frac{\ 2\pi\ }{\ \lambda\ }\,}
	I_{\nu+n}\left(\frac{\ 1\ }{\ \lambda\ }\right)
	\exp\left(-\frac{\ 1\ }{\ \lambda\ }-
	\frac{\ \lambda\ }{\ 8\ }\right).
\end{equation}
We can therefore rewrite the short time kernel as
\begin{equation}
\begin{aligned}
	{\mathcal K}^{(\nu)}(\theta,\theta{'};\epsilon)\sim&
	\frac{\ 2^{2\nu}\{\varGamma(\nu)\}^{2}\ }{\ \sqrt{2\pi\lambda\,}\ }
	(\sin\theta\sin\theta{'})^{\nu}
	\exp\left(-\frac{\ 1\ }{\ \lambda\ }-
	\frac{\ \lambda\ }{\ 8\ }\right)\\
	&\times
	\sum_{n=0}^{\infty}\frac{\ n!(\nu+n)\ }{\ \varGamma(2\nu+n)\ }
	I_{\nu+n}\left(\frac{\ 1\ }{\ \lambda\ }\right)
	C_{n}^{\nu}(\cos\theta)C_{n}^{\nu}(\cos\theta{'}).
\end{aligned}
\end{equation}
The sum in the right hand side above can be converted into a simplified form by the formula(see e.g. Ch. 11.5 of ref.\cite{Watson} or Ch. 8.8 of ref.\cite{KleinertBook})
\begin{equation}
\begin{aligned}
	&\frac{\ 2^{2\nu}\{\varGamma(\nu)\}^{2}\ }{\ \sqrt{2\pi\lambda\,}\ }
	(\sin\theta\sin\theta{'})^{\nu}
	\sum_{n=0}^{\infty}\frac{\ n!(\nu+n)\ }{\ \varGamma(2\nu+n)\ }
	I_{\nu+n}\left(\frac{\ 1\ }{\ \lambda\ }\right)
	C_{n}^{\nu}(\cos\theta)C_{n}^{\nu}(\cos\theta{'})\\
	&=\frac{\ (\sin\theta\sin\theta{'})^{1/2}\ }{\lambda}
	\exp\left({\frac{\ \cos\theta\cos\theta{'}\ }{\lambda}}\right)
	I_{\nu-1/2}\left(\frac{\ \sin\theta\sin\theta{'}}{\lambda}\right)
\end{aligned}
\end{equation}
to result in
\begin{equation}
\label{eq:kernel02}
	{\mathcal K}^{(\nu)}(\theta,\theta{'};\epsilon)\sim
	\frac{\ (\sin\theta\sin\theta{'})^{1/2}\ }{\lambda}
	\exp\left({
	-\frac{\ 1-\cos\theta\cos\theta{'}\ }{\lambda}-
	\frac{\ \lambda\ }{\ 8\ }}\right)
	I_{\nu-1/2}\left(\frac{\ \sin\theta\sin\theta{'}}{\lambda}\right).
\end{equation}
We have thus obtained a closed expression for the short time kernel for arbitrary values of $\nu\ge1/2$.

To check the validity of \eqref{eq:kernel02}, let us first set $\nu$ to be unity and consider the case of the free particle in a box.
Since $I_{1/2}(z)$ can be expressed as $I_{1/2}(z)=\sqrt{2/(\pi z)\,}\sinh z$,
the modified Bessel function in \eqref{eq:kernel02} yields
\begin{equation}
\label{eq:modBessel01}
	\sqrt{\frac{\lambda}{\ 2\pi\sin\theta\sin\theta{'}\ }\,}\left\{
	\exp\left(\frac{\ \sin\theta\sin\theta{'}\ }{\lambda}\right)-
	\exp\left(-\frac{\ \sin\theta\sin\theta{'}\ }{\lambda}\right)\right\}
\end{equation}
for $\nu=1$. Hence the right hand side of \eqref{eq:kernel02} can now be rewritten as
\begin{equation}
\label{eq:kernel03}
	\frac{e^{\lambda/8}}{\ \sqrt{2\pi\lambda\ }\,}\left[
	\exp\left\{-\frac{\ 1-\cos(\theta-\theta{'})\ }{\lambda}\right\}-
	\exp\left\{-\frac{\ 1-\cos(\theta+\theta{'})\ }{\lambda}\right\}\right].
\end{equation}
This kernel possesses infinitely many saddle points to make us replace it with
\begin{equation}
\label{eq:kernel04}
	\frac{1}{\ \sqrt{2\pi\lambda\ }\,}\sum_{k=-\infty}^{\infty}\left[
	\exp\left\{-\frac{\ (\theta-\theta{'}-2k\pi)^{2}\ }{2\lambda}\right\}-
	\exp\left\{-\frac{\ (\theta+\theta{'}-2k\pi)^{2}\ }{2\lambda}\right\}\right]
\end{equation}
for infinitesimally small $\lambda$. Here use has been made of the method given by the present author in ref.\cite{Sakoda2017} for converting complex kinetic term in the path integral into the stand one with additional potential terms. For the present case the additional potential is same for all saddle points and given by $-\lambda/8$ in the exponent of the Feynman kernel to cancel the factor $e^{\lambda/8}$ in \eqref{eq:kernel03}.
We now observe that \eqref{eq:kernel04} is nothing but the Feynman kernel, that is normalized to fit integration with respect to $\theta$($\theta{'}$) instead of $x$($x{'}$), for the free particle in a box. Therefore our derivation of the short time kernel \eqref{eq:kernel02} is correct for $\nu=1$. It should be emphasized here that, in the calculation above, the origin of the minus sign in front of the second term in the sum in \eqref{eq:kernel04} is the coefficient of $e^{-z}$ in $\sinh z$ appears in $I_{1/2}(z)=\sqrt{2/(\pi z)\,}\sinh z$.

For $\nu=2$, we make use of
$I_{3/2}(z)=\sqrt{2/(\pi z)\,}(\cosh z-z^{-1}\sinh z)$ to obtain
\begin{equation}
\begin{aligned}
	&\left(1-\frac{\lambda}{\ \sin\theta\sin\theta{'}\ }\right)
	\exp\left(\frac{\ \sin\theta\sin\theta{'}\ }{\lambda}\right)+
	\left(1+\frac{\lambda}{\ \sin\theta\sin\theta{'}\ }\right)
	\exp\left(-\frac{\ \sin\theta\sin\theta{'}\ }{\lambda}\right)\\
	=&
	\exp\left(\frac{\ \sin\theta\sin\theta{'}\ }{\lambda}-
	\frac{\lambda}{\ \sin\theta\sin\theta{'}\ }\right)+
	\exp\left(-\frac{\ \sin\theta\sin\theta{'}\ }{\lambda}+
	\frac{\lambda}{\ \sin\theta\sin\theta{'}\ }\right)
\end{aligned}
\end{equation}
by discarding irrelevant terms, with the same pre-factor as in \eqref{eq:modBessel01} for the modified Bessel function in \eqref{eq:kernel02}.
We then find that the short time kernel can be written as
\begin{equation}
\label{eq:kernel05}
\begin{aligned}
	{\mathcal K}^{(2)}(\theta,\theta{'};\epsilon)
	=&
	\frac{1}{\ \sqrt{2\pi\lambda\ }\,}\sum_{k=-\infty}^{\infty}\left[
	\exp\left\{-\frac{\ (\theta-\theta{'}-2k\pi)^{2}\ }{2\lambda}-
	\frac{\lambda}{\ \sin\theta\sin\theta{'}\ }\right\}\right.\\
	&\hphantom{\frac{1}{\ \sqrt{2\pi\lambda\ }\,}\sum_{k=-\infty}^{\infty}}
	\!\!\!+\left.
	\exp\left\{-\frac{\ (\theta+\theta{'}-2k\pi)^{2}\ }{2\lambda}+
	\frac{\lambda}{\ \sin\theta\sin\theta{'}\ }\right\}\right].
\end{aligned}
\end{equation}
This is the decomposition of the short time kernel for $\nu=2$ into the sum over paths with different number of reflections. The second term in the sum above expresses contributions from paths reflected odd times at boundaries. Surprisingly, the coefficient of this terms is now $+1$.
For integral values of $\nu$, we can repeat the similar prosess to find that the coefficient is $-1$ if $\nu$ is odd integer and $+1$ for even integers.

We now proceed to consider the case of non-integral values for $\nu$. For this case we may resort to making use of the asymptotic form of the modified Bessel function in \eqref{eq:kernel02}. To take all possible contributions from reflected paths into account, we have to determine the $\arg(\sin\theta\sin\theta{'})$. In the original domain, both $\sin\theta$ and $\sin\theta{'}$ are positive real. We thus define $\arg(\sin\theta\sin\theta{'})=0$ there. It is then natural to define $\arg(\sin\theta\sin\theta{'})=-\pi$ for the saddle point at $\theta+\theta{'}=0$, $\arg(\sin\theta\sin\theta{'})=-2\pi$ for $\theta-\theta{'}=-2\pi,\,\ldots$ and $\arg(\sin\theta\sin\theta{'})=\pi$ for the saddle point at $\theta+\theta{'}=2\pi$, $\arg(\sin\theta\sin\theta{'})=2\pi$ for $\theta-\theta{'}=2\pi,\,\ldots$. In this way, we obtain, by keeping only relevant terms, the asymptotic form of the modified Bessel function in \eqref{eq:kernel02} for the saddle point at $\theta-\theta{'}=2k\pi$($k=0,\,\pm1,\,\pm2,\,\ldots$)
\begin{equation}
	e^{2k\nu\pi i}
	\exp\left\{\frac{\ \sin\theta\sin\theta{'}}{\lambda}-
	\frac{\ \lambda\ }{\ 2\ }
	\frac{\ \nu(\nu-1)\ }{\ \sin\theta\sin\theta{'}\ }\right\}
\end{equation}
and
\begin{equation}
	e^{(2k-1)\nu\pi i}
	\exp\left\{-\frac{\ \sin\theta\sin\theta{'}}{\lambda}+
	\frac{\ \lambda\ }{\ 2\ }
	\frac{\ \nu(\nu-1)\ }{\ \sin\theta\sin\theta{'}\ }\right\}
\end{equation}
for saddle point at $\theta+\theta{'}=2k\pi$, with the same pre-factor that appears in \eqref{eq:modBessel01}.
We thus obtain
\begin{equation}
	\label{eq:kernel06}
\begin{aligned}
	{\mathcal K}^{(\nu)}(\theta,\theta{'};\epsilon)
	=&
	\frac{1}{\ \sqrt{2\pi\lambda\ }\,}\sum_{k=-\infty}^{\infty}\left[
	e^{2k\nu\pi i}
	\exp\left\{-\frac{\ (\theta-\theta{'}-2k\pi)^{2}\ }{2\lambda}-
	\frac{\ \lambda\ }{\ 2\ }
	\frac{\nu(\nu-1)}{\ \sin\theta\sin\theta{'}\ }\right\}\right.\\
	&\hphantom{\frac{1}{\ \sqrt{2\pi\lambda\ }\,}\sum_{k=-\infty}^{\infty}}
	\!\!\!\!\!\!\!\!\!\!\!\!+\left.
	e^{(2k-1)\nu\pi i}
	\exp\left\{-\frac{\ (\theta+\theta{'}-2k\pi)^{2}\ }{2\lambda}+
	\frac{\ \lambda\ }{\ 2\ }
	\frac{\nu(\nu-1)}{\ \sin\theta\sin\theta{'}\ }\right\}\right]
\end{aligned}
\end{equation}
as the decomposition of the Feynman kernel \eqref{eq:kernel01} into the sum over paths with difference in the number of reflections for infinitesimally small $\epsilon$. The phase factor in front of each component above may change if we choose another prescription to determine $\arg(\sin\theta\sin\theta{'})$ outside the original domain. We may choose such that $\arg(\sin\theta\sin\theta{'})=0$ for $\theta$ in $2k\pi<\theta<(2k+1)\pi$ and $\arg(\sin\theta\sin\theta{'})=\pi$ for $(2k-1)\pi<\theta<2k\pi$ for example. For this choice all coefficients in the sum of contributions from paths reflected even times become unity and those in the sum of contributions from paths reflected odd times reduce to $e^{\nu\pi i}$. We therefore observe here that the factor $-1=e^{\pi i}$ for the reflection of the free particle in a box is not the universal one; it rather depends on the parameter $\nu$ that characterizes the potential.

\section{Summary}
We have studied the decomposition of the Feynman kernel for a particle in a box with $1/\sin^{2}\theta$ potential to find that the phase the kernel acquires at each reflection by boundaries depends on the parameter of the potential. The form of the decomposition possesses Lagrangian form of the Euclidean action and allows us to consider that the Feynman kernel can be expressed as a sum over paths if we treat the phase generated by reflection at boundaries carefully. The phase which appears in Euclidean path integral may have some geometric origin. It will be, therefore, interesting to find its meaning. Finally, we must add the following comment; although we have obtained our result starting from the eigenfunction expansion of the Feynman kernel, it will be desired to find a method to arrive the same result from the Hamiltonian path integral, as we usually do in obtaining the Lagrangian path integral for systems on the whole real line, by keeping good connection with the operator formalism so that we can deduce the eigenfunction expansion of the kernel solely by means of the path integral technique. Such a method will be reported elsewhere\cite{Sakoda2018a}.

\end{document}